\documentclass[letterpaper,10pt,oneside,onecolumn]{article}
\usepackage[left=4cm, right=4cm, top=2cm]{geometry}
\usepackage{epsfig,amsbsy,bm}
\usepackage{etoolbox}
\usepackage{color}
\usepackage{placeins}
\usepackage{mathrsfs}
\usepackage{amsmath}
\usepackage{graphicx}
\usepackage{float}
\usepackage{dcolumn}
\usepackage[nice]{nicefrac}
\usepackage[tight]{units}
\usepackage{lipsum}
\usepackage[dvipsnames]{xcolor}
\usepackage{pgfplots}
\usepackage[utf8]{inputenc}
\usetikzlibrary{patterns}
\usetikzlibrary{arrows.meta}
\usetikzlibrary{fadings}
\usetikzlibrary{spy}
\usepackage{gensymb}
\tikzset{>={Latex[width=2mm,length=2mm]}}
\pgfplotsset{compat=newest}
\pgfplotsset{plot coordinates/math parser=false}
\usepgfplotslibrary{colorbrewer}
\pgfplotsset{
    colormap/Spectral
}

\pgfplotscolorbarCMYKworkaroundfalse

\def\t0{\mbox{$t_{\mbox{{\tiny {0}}}}$}}

\def\p0{\mbox{$p_{\mbox{{\tiny {0}}}}$}}

\def\E0{\mbox{$E_{\mbox{{\tiny {0}}}}$}}

\newcommand{\vect}{\vec}

\newcommand{\be}{\begin{equation}}
\newcommand{\ee}{\end{equation}}
\newcommand{\ba}{\begin{eqnarray}}
\newcommand{\ea}{\end{eqnarray}}

\begin{document}

\title{Temporal Resolution of the {\it RABBITT} technique}

\author{M. Isinger, D. Busto, S. Mikaelsson, S. Zhong, C. Guo, P. Sali\`eres,\\ C. L. Arnold, A. L'Huillier and M. Gisselbrecht}



\maketitle

\begin{abstract}
One of the most ubiquitous techniques within attosecond science is the so-called Reconstruction of Attosecond Bursts by Interference of Two-Photon Transitions ({\it RABBITT}). Originally proposed for the characterization of attosecond pulses, it has been successfully applied to accurate determinations of time delays in photoemission. Here, we examine in detail, using numerical simulations, the effect of the spatial and temporal properties of the light fields and of the experimental procedure on the accuracy of the method. This allows us to identify the necessary conditions to achieve the best temporal resolution in {\it RABBITT} measurements.
\end{abstract}

\section{Introduction}
The study of ultrafast dynamics of small quantum systems is currently undergoing a rapid progression due to the emergence of intense and ultra-short pulsed light sources in the eXtreme UltraViolet (XUV) and X-rays regimes. One driving force behind this progress is the field of attosecond science,
thanks to light sources based on High-order Harmonic Generation (HHG) in gases, allowing the investigation of electron dynamics at the fastest available timescale \cite{KrauszRMP2009}. A world-spanning effort has led to the generation of pulses with duration below 100 as \cite{SansoneScience2006,GoulielmakisScience2008,LiNC2017,
GaumnitzOE2017}, photon energies up to several keV \cite{PopmintchevScience2012,CousinPRX2017} and pulse energies up to a few $\mu$J \cite{LeDeroffPRA2000,MidorikawaPiQE2008}, enabling the exploration of real-time dynamics of fundamental processes induced by light matter-interaction, such as photoionization~\cite{IsingerScience2017}. To push further the frontier of attosecond science and reach the ultimate temporal resolution, the metrology of the available experimental techniques needs to be addressed.

Two major techniques, commonly referred to as {\it Streaking} \cite{ConstantPRA1997,HentschelN2001} and {\it RABBITT} \cite{VeniardPRA1996, PaulScience2001,MairesseScience2003}, are currently used to get an insight into the attosecond dynamics of photoemission. Both techniques are based on cross-correlation measurements between the attosecond XUV light field and a phase locked infrared (IR) laser field. The measurements consist in producing electron wave-packets, whose temporal characteristics essentially mimic those of the ionizing XUV light field. The electron wave-packets are thereafter probed, in the vicinity of the ionic core, by an IR field of intensity varying from 10$^{11}$ Wcm$^{-2}$ to 10$^{13}$ Wcm$^{-2}$. Theoretical works have provided a deep understanding of the temporal properties of these electron wave packets, allowing to disentangle the contributions of the light fields and of the photoionization dynamics \cite{DahlstromCP2013,PazourekRMP2015}. However, from an experimental perspective, the temporal information that can be extracted from the electron wave-packets is often limited by the precise knowledge (or control) of the light properties, including temporal and spatial overlap, and the convergence of the numerical algorithms used to analyze the data.

{\it Streaking} utilizes a single attosecond pulse and an intense IR field to streak the photoelectrons at different times. Several rigorous studies have discussed its robustness \cite{GagnonOE2009}, the effect of the laser jitter \cite{ZhongOE2013} and lately the reconstruction method \cite{GaumnitzOE2017}. {\it RABBITT}, which stands for Reconstruction of Attosecond Beating by Interference of Two-photon Transition, relies on an attosecond pulse train and a weak probe field to induce interference between multiple quantum paths in the spectral domain. The {\it RABBITT} technique has been compared to {\it Streaking} in an extensive study \cite{CattaneoOE2016}, as well as to two-photon volume autocorrelation \cite{KrusePRA2010}. There is, however, to our knowledge, neither a detailed investigation of the problems inherent to {\it RABBITT} measurements, nor a discussion of the ultimate time resolution that can be reached.

In this work, we present a comprehensive analysis of the {\it RABBITT} technique. We present its principle in section 2 and investigate the influence of the properties of the light fields in section 3. We discuss the statistical effects inherent to any measurement in order to define optimal experimental parameters in section 4.

\section{Theory and Experiment}

\subsection{Principle}

The principle of the {\it RABBITT} technique is presented in figure \ref{rabbit}. An XUV attosecond pulse train (APT), corresponding in the frequency domain to a comb of odd-order harmonics, is overlapped spatially and temporally with an IR pulse, which is often a weak replica of the field used to generate the APT. The interaction between atoms or molecules with both light fields leads to the creation of electron wave-packets, composed of peaks due to ionization by the XUV attosecond pulse train and sidebands, which arise from two-photon transitions (XUV$\pm$IR). The sideband signals oscillate as the delay between the IR and XUV pulses is varied. This is due to an interference between two quantum paths, the first one involving absorption of one harmonic and an IR photon, and the second involving absorption of the next harmonic and emission of an IR photon. We use second-order perturbation theory for the light-matter interaction, expressed with the dipole approximation, make the rotating-wave approximation, considering only the processes where the XUV fields are absorbed first, and assume that the fundamental and harmonic fields are monochromatic (The latter assumption will be removed in Section 3.2). For a single atom/molecule the $q$\textsuperscript{th} sideband signal can be described by the equation \cite{VeniardPRA1996},
\begin{equation}
S_q=\lvert\mathcal{A}_\mathrm{abs}+ \mathcal{A}_\mathrm{emi}\rvert^2=\lvert\mathcal{A}_\mathrm{abs}\rvert^2+ \lvert\mathcal{A}_\mathrm{emi}\rvert^2+ 2\lvert\mathcal{A}_\mathrm{abs}\rvert\lvert\mathcal{A}_\mathrm{emi}\rvert\cos\left[\arg(\mathcal{A}_\mathrm{abs})-\arg(\mathcal{A}_\mathrm{emi})\right],
\label{rabbit}
\end{equation}
where $\mathcal{A}_\mathrm{abs}=\mathcal{E}_1\mathcal{E}_{q-1}M_\mathrm{abs}$ and $   \mathcal{A}_\mathrm{emi}=\mathcal{E}^*_1\mathcal{E}_{q+1}M_\mathrm{emi}$. In these expressions,
$\mathcal{E}_1$ and $\mathcal{E}_{q\pm 1}$ denote the amplitudes of the fundamental and $(q\pm 1)$\textsuperscript{th} harmonic fields, while $M_\mathrm{abs}$ ($M_\mathrm{emi}$) are two-photon ionization transition amplitudes involving absorption (emission) of the fundamental field in the continuum. It is often assumed that the phase of the fundamental field is constant, and simply related to that of the attosecond pulse train by a variable delay $\tau$, so that $\arg(\mathcal{E}_1)-\arg(\mathcal{E}^*_1)=2\omega\tau$, \mbox{$\omega$} denoting the fundamental frequency ($\pi/\omega=1.3$ fs at 800 nm). We further denote $\arg(\mathcal{E}_{q+1})-\arg(\mathcal{E}_{q-1})=2\omega\tau_\textsc{XUV}$ and $\arg(M_\mathrm{emi})-\arg(M_\mathrm{abs})=2\omega\tau_\mathrm{A}$, so that Eq.~\ref{rabbit} becomes
\begin{equation}
  S_q=\lvert\mathcal{A}_\mathrm{abs}\rvert^2+ \lvert\mathcal{A}_\mathrm{emi}\rvert^2+ 2\lvert\mathcal{A}_\mathrm{abs}\rvert\lvert\mathcal{A}_\mathrm{emi}\rvert\cos[2\omega (\tau-\tau_\mathrm{R})],
  \label{rabbit2}
\end{equation}
where  $\tau_\mathrm{R}=\tau_{\textsc{XUV}} + \tau_\mathrm{A}$. The first term is the group delay of the attosecond pulses in the train, which is here assumed to be perfectly periodic, and the second term comes from the phase difference between the two-photon matrix elements \cite{PaulScience2001,DahlstromCP2013}. An accurate determination of $\tau_\mathrm{R}$ requires good control over the delay $\tau$. In the experiments, the delay $\tau_\mathrm{R}$ is obtained through an analysis of {\it RABBITT} spectrograms [see Fig.~\ref{rabbit}(b)], which implies a volume integration over the interaction region, according to
\begin{equation}
    S_q=\int \left( \lvert\mathcal{A}_\mathrm{abs}\rvert^2+ \lvert\mathcal{A}_\mathrm{emi}\rvert^2+ 2\lvert\mathcal{A}_\mathrm{abs}\rvert\lvert\mathcal{A}_\mathrm{emi}\rvert\cos[2\omega (\tau-\tau_\mathrm{R})]\right)d^3\vect{r},
  \label{rabbit3}
\end{equation}
The dependence of the amplitudes $\mathcal{A}_\mathrm{abs}$ and $\mathcal{A}_\mathrm{emi}$ on space do not affect the precision of the {\it RABBITT} measurement. However, when $\tau_\mathrm{R}$, i.e. $\tau_\textsc{XUV}$, depends on space over the region of overlap, the sideband oscillations become blurred, leading to an inaccuracy in the phase retrieval. One aim of this article is to study the limitations to the measurement induced by such spatial variations.

\begin{figure}[]
   \def\svgwidth{8cm}
   \includegraphics{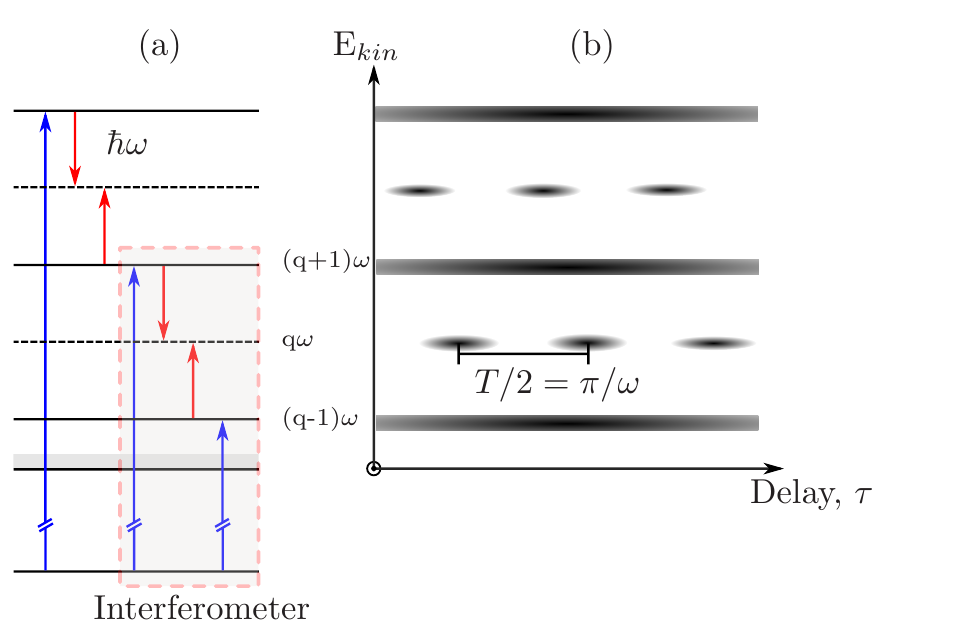}
   \centering
     \caption{Principle of RABBITT. Each sideband $q$ makes up a quantum interferometer, one arm being the absorption of harmonic $q-1$ followed absorption of an IR photon, and the other the absorption of harmonic $q+1$ followed by emission of an IR photon, leading to the same final energy. The intensity of the sideband oscillates as a function of delay between the pulses, with an oscillation period of T/2 where T is the period of the fundamental, and the phase of this oscillation reveals information about the target system.}
  \label{rabbit}
\end{figure}

 To eliminate the influence of the excitation pulse, two measurements can be performed simultaneously, for example, on different ionization processes \cite{SchulzeScience2010,KlunderPRL2011,ManssonNP2014,HeuserPRA2016,OssianderNatPhys2016, IsingerScience2017} or in different target species \cite{PalatchiJPB2014,GuenotJPB2014,SabbarPRL2015,JordanPRA2017}. This enables the determination of a relative photoionization time delay $\Delta \tau_\mathrm{A}$. The absolute photoionization delay $\tau_\mathrm{A}$ can be deduced if we assume that one of the delays can be sufficiently accurately calculated to serve as an absolute reference  \cite{OssianderNatPhys2016}.

Originally, the {\it RABBITT} technique has consisted in analyzing the energy-integrated oscillations of the sideband signal. State-of-the-art photoelectron spectrometers provide now high spectral resolution which allows for an {\it energy-resolved} analysis of the sideband oscillations. This technique, dubbed {\it Rainbow RABBITT} \cite{GrusonScience2016}, can be used to retrieve the amplitude and phase variation across sidebands, which is useful to disentangle contributions from different ionization processes contributing to the same sideband \cite{GuenotPRA2012,IsingerScience2017}, or to characterize complex photoelectron wavepackets \cite{GrusonScience2016,BeaulieuScience2017,BustoJPB2018}.

\subsection{Experimental Methods}

Various experimental set-ups have been designed for the {\it RABBITT} technique \cite{KlunderPRL2011,MullerOL2001,PaulScience2001}. The common approach is to use a Mach-Zehnder-type interferometer \cite{SansoneS2006,ChiniOE2009} to control the delay between the pump and the probe beam. The interferometer is said to be passively stabilized if a movement in one arm is compensated for in the other arm by a clever positioning of the mirrors. On the other hand, it is actively stabilized if the delay stage is controlled by a feedback loop, actively monitoring the delay and accounting for drifts \cite{KroonOL2014}. This ensures both short term (attosecond to femtosecond) and long term (hours to days) stability of the delay between the two arms\cite{ChiniOE2009,HuppertRSI2015}. However, the jitter between the pump and probe arm cannot be fully corrected and typical values of this delay jitter ranges from 200 as RMS (passive stabilization) down to 50 as RMS (active stabilization). With the recent implementation of a beam pointing stabilization technique to account for any spatial drifts that occur on a rate of 100 Hz or slower, we can currently reach delay jitters on the order of 25 as. Efforts are made to improve these values even further.

Once the experiment is carried out, the spectrogram needs to be post-processed in order to extract the phase of individual sidebands. One approach is to perform Fourier analysis to efficiently filter away low and high frequency noise and then analyze the traces using a robust non-linear Levenberg-Marquardt\cite{LevenbergQAM1944, MarquardtJSIAM1963} algorithm. The algorithm returns the best-fit phase as well as an estimate of the fit goodness, essentially given by the statistical properties of the signal.

\subsection{Simulations}

\begin{figure}[htbp]
\centering
   \includegraphics{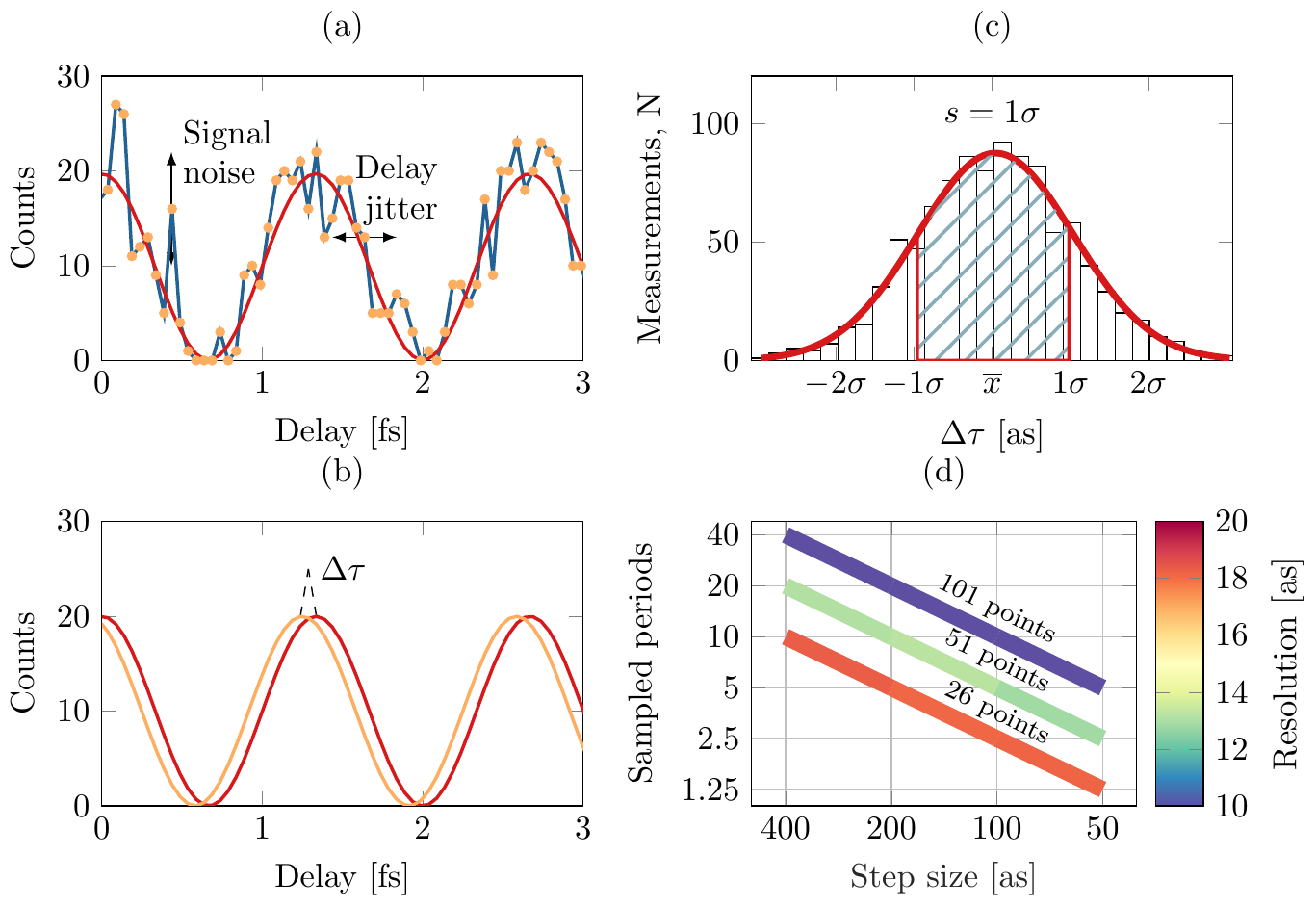}
     \caption{Illustration of the model used to simulate {\it RABBITT} traces. (a) A {\it RABBITT} oscillation, on which temporal jitter and signal fluctuation are added; (b) Determination of the phase difference between the best fit and the original function; (c) After repetition of this procedure $N$ times, the resolution can be determined as the standard deviation of the distribution function; (d) Resolution (in color) as a function of number of sampled periods and step size for different number of samples.}
  \label{methodstep}
\end{figure}

Finally, we present how we model {\it RABBITT} measurements, using Monte-Carlo simulations. A sample is obtained by discretizing Eq. \eqref{rabbit2} for a certain number of periods and a given step size, as seen in figure \ref{methodstep}(a). Each sample is then perturbed by an insertion of noise in both dimensions, to emulate interferometer instability and statistical noise. The temporal jitter is drawn from a normal distribution with a variance $\sigma_t^2$ of a pre-set value. The jitter of the electron signal is drawn from a Poisson distribution with a variance $\sigma_s^2$, equal to the acquired counts at each step. The noisy signal is then used as input for the non-linear least square fit [figure~\ref{methodstep}(b)], and the phase difference between the best fit and the original unperturbed cosine is evaluated giving the error of the fitted phase. This process is repeated $N$ times to acquire enough statistics to reveal a normal distribution of values around the mean value $\overline{x} = \sum ^{N}_{i=1}x_{i}/N$, as shown in figure~\ref{methodstep}(c). The normal distribution is fitted and the resolution, defined as the standard deviation $s$, can be obtained as
\begin{equation}
s = \sqrt{\frac{1}{N-1} \sum_{i=1}^N (x_i - \overline{x})^2}
\label{std}
\end{equation}

Figure~\ref{methodstep}(d) shows how the resolution varies as a function of sampled periods and step size. The step size is decreasing as the number of sampled periods is decreasing, so as to keep the number of sample points constant. It is of great importance to realize that it is the total number of sample points that determine the temporal resolution with this phase retrieval algorithm, as long as the oscillation frequency is resolved.

Note that the resolution should not to be confused with the standard deviation of the mean (often called standard error of mean or SEM), which would be a value of how close the estimated mean from a set of $N$ {\it RABBITT} measurements is to the true mean, and scales as $s/\sqrt{N}$. To approach the true mean, we can either minimize the error, $s$, of each measurement or acquire a larger set, $N$, of data.

\section{Influence of the spatial and temporal properties of the light fields}

In this section, we investigate the influence of the properties of the harmonic radiation on the resolution of the extracted time delay from the {\it RABBITT} technique.

\subsection{Spatial variations of the {\it RABBITT} phase}

The $(q\pm 1)$\textsuperscript{th} harmonic field has a phase equal to $(q\pm 1)\phi_\mathrm{fund} + \Phi_{q\pm 1}$, where $\phi_\mathrm{fund}$ is the phase of the fundamental field and $\Phi_{q\pm 1}$ is the dipole phase, accumulated by the electron on its trajectory. For a Gaussian beam, the spatial phase is $\phi(r,z)=-kz-kr^2/R(z)+\zeta(z)$, where $R(z)=z(1+z_0^2/z^2)$ is the radius of curvature and $\zeta(z)=\arctan(z/z_0)$ the Gouy phase ($z_0$ is the Rayleigh length). The spatially-dependent phase entering the {\it RABBITT} equation contains two terms
\begin{eqnarray}
       &2\arg(\mathcal{E}_1)=2\phi_\mathrm{probe}, \\
&\arg(\mathcal{E}_{q+1})-\arg(\mathcal{E}_{q-1})=2\phi_\mathrm{fund}+ \Delta\Phi_{q}.
\end{eqnarray}
where $\Delta\Phi_q=\Phi_{q+1}-\Phi_{q-1}$.

Let us assume that the probe field and fundamental field generating the harmonics have the same phase.
This requires that the same focusing optics is used in both probe and pump arms. Note that imperfect focusing of the attosecond light may induce a spreading of the radiation both in space and in time due to optical aberrations \cite{BouchetOE2011, BouchetOE2013, BouchetJOSAA2010}, leading to distorsions of the corresponding RABBIT traces. In the following, we do not consider such effects.
To include the influence of the dipole phase, we use an analytic expression obtained by linearizing the variation of the emitted frequency with return time for each of the trajectories contributing to the harmonic emission \cite{GuoJPBAMOP2018,WikmarkPNAS2018}. For the short trajectory used in {\it RABBITT} measurements,
\begin{equation}
\Phi_{q\pm 1}= \frac{\gamma}{I}(q\pm 1-q_p)^2\omega^2.
\label{eq:phasef}
\end{equation}
where $\gamma$ is a gas independent coefficient equal to $1.03\times 10^{-18}$ s\textsuperscript{2}~W~cm\textsuperscript{-2} at 800~nm, $q_p=I_p/\omega$ where $I_p$ is the ionization potential, and $I$ is the laser intensity. The difference of phase between the two quantum paths entering Eq.~(\ref{rabbit2}) due to the harmonic fields is therefore
\begin{equation}
\Delta \Phi_{q}= \frac{4\gamma}{I}(q-q_p)\omega^2.
\label{eq:phasediff}
\end{equation}
This expression gives a simple estimate of the group delay of the attosecond pulses, also called ``attochirp".
\begin{equation}
\tau_\textsc{XUV}= \frac{\Delta \Phi_{q}}{2\omega}=\frac{2\gamma}{I}(q-q_p)\omega,
\label{tau}
\end{equation}
which varies linearly with the harmonic order, except close to the cutoff frequency  \cite{MairesseScience2003,KazamiasPRA2004}.

\begin{figure}[]
\centering
   \includegraphics{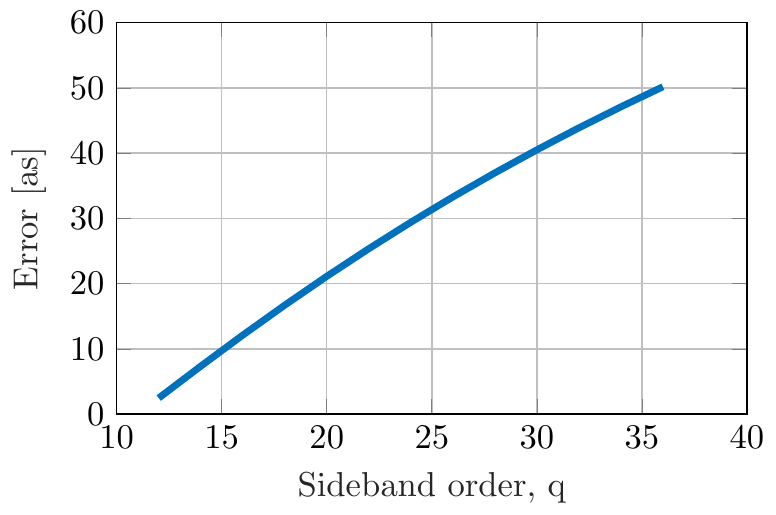}
     \caption{Error due to spatial integration as a function of harmonic order for a beam waist of 25 microns.}
  \label{fig:t_and_r}
\end{figure}

This expression also shows that $\tau_\textsc{XUV}$ depends on the radial coordinate, $r$, through the variation of the laser intensity $I$.
We can now estimate the influence of wavefront mismatch due to the dipole phase on the accuracy of the {\it RABBITT} measurements.
We assume that $I=I_0 \exp(-2r^2/w^2)$, where $I_0$ is the laser peak intensity at focus and $w$ the radius. The phase difference in a {\it RABBITT} measurement can therefore be expressed as
\begin{equation}
\Delta\Phi_{q}(r) = \frac{4\gamma(q-q_p)\omega^2}{I_0}\exp\left(\frac{2r^2}{w^2}\right).
\end{equation}
At $r=0$, this quantity corresponds to the attosecond chirp\cite{MairesseScience2003}. We insert this phase variation in Eq.~(\ref{rabbit3}) and calculate its influence on the volume-integrated {\it RABBITT} traces. Since the results are relatively independent of the focusing condition, we only present the error added by the wavefront mismatch due to the dipole phase at a given beam waist in Figure~\ref{fig:t_and_r}. As expected, the error increases with the harmonic order and will add a small offset to the attosecond chirp which is usually about one order of magnitude higher. This error can be eliminated by performing two experiments simultaneously.

This result explains why the spatial dependence of the phase of the fields in Eq.~(\ref{rabbit}) has not been discussed in {\it RABBITT} measurements, though the importance of wavefront matching of the pump and probe beams has been pointed out \cite{LopezMartensPRL2005}. If the focusing geometries for the probe and harmonic beams are, however, different, not only errors due to wavefront mismatch may appear, but also the terms including the Gouy phase do not necessarily cancel out and need also to be considered in the error budget. In addition, possible delays between the harmonic fields and the probe field on the way to the application chamber, due to reflections on focusing mirrors for example, should also be taken into account \cite{DinuPRL2003,LocherOptica2015}.

\subsection{Influence of the temporal properties of the light fields}

The dipole phase has two effects on the temporal properties of the XUV radiation. It leads to a postive group delay of the attosecond pulses [see Eq.~\ref{tau}]. It also leads to a femtosecond chirp of the individual harmonics (in the frequency domain, to group delay dispersion) \cite{VarjuJoMO2005}.

\begin{figure}[]
    \centering
   \includegraphics{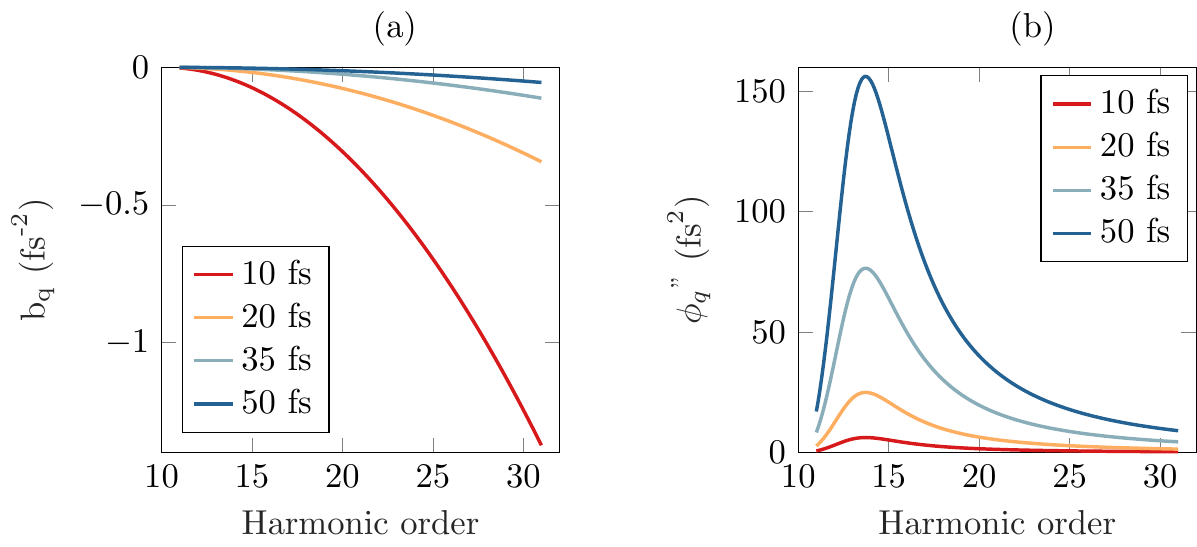}
    \caption{Intrinsic femtosecond chirp of the XUV radiation. Chirp-rate a) and Group Delay Dispersion b) of individual harmonic for different pulse durations of the fundamental laser.}
 \label{fig:Femtochirp}
\end{figure}

For a Gaussian pulse with intensity profile $I(t)=I_0\exp[-at^2/\tau^2]$, where $a=4\ln(2)$ and $\tau$ is the pulse duration (full width at half maximum), Eq.~(\ref{eq:phasef}) can be approximated by:
\begin{equation}
\Phi_{q\pm 1}\approx \frac{\gamma (q\pm 1-q_p)^2\omega^2}{I_0}\left(1+\frac{at^2}{\tau^2}\right).
\end{equation}
The  chirp coefficient, so-called chirp-rate\cite{VarjuJoMO2005} is,
\begin{equation}
b_{q\pm 1} \approx -\frac{a\gamma (q\pm 1-q_p)^2\omega^2}{\tau^2I_0}= -\frac{a\gamma (q\pm 1-q_p)^2(2\pi)^2}{T^2\tau^2I_0},
\label{eq:chirprate}
\end{equation}
$T$ denoting the laser cycle. The corresponding group delay dispersion (GDD), $\phi_{q\pm 1}^{''}$, is related to the chirp-rate by
\begin{equation}
\phi_{q\pm 1}^{''}=-\frac{b_{q\pm 1}\tau_{q\pm 1}^4}{a^2+b_{q\pm 1}^2\tau_{q\pm 1}^4},
\end{equation}
where $\tau_{q\pm 1}\sim \tau/2$ is the full width at half maximum pulse duration of the $(q\pm 1)$\textsuperscript{th} harmonic\cite{DurfeePRL1999}, also assumed to be Gaussian. The chirp coefficient and the corresponding group delay dispersion for different harmonics order are evaluated using an analytic model for high harmonic generation as a function of the fundamental laser pulse duration\cite{GuoJPBAMOP2018}.

\begin{figure}[]
\centering
   \includegraphics{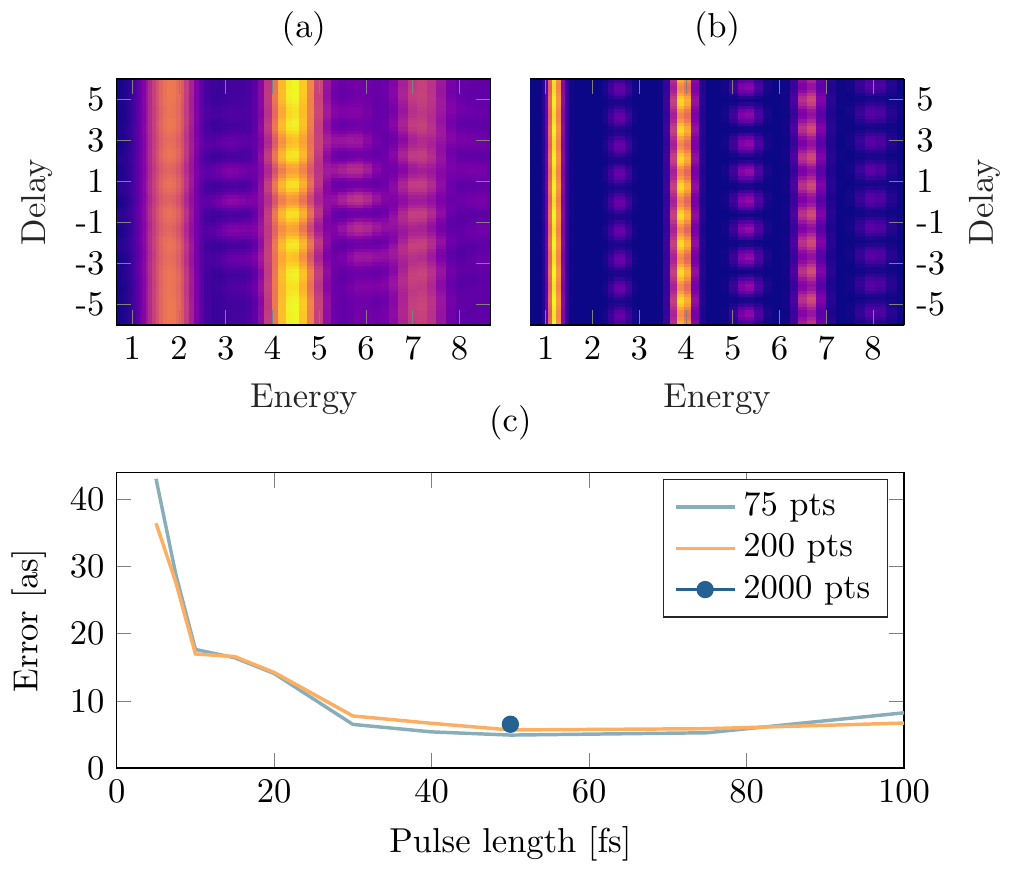}
  \caption{Effect of the chirp-rate as function of the fundamental pulse duration in {\it RABBITT} measurement. Simulated scans in Helium are exhibited for pulses of 5 fs and 30 fs, panels a) and b) respectively. The error in the phase retrieval for sideband 19 (at 5-6 eV kinetic energy) is shown in panel c). To ensure the convergence of the results, simulations were performed by varying the number of sampling points.}
  \label{fig:SFA_chirprate}
\end{figure}

\begin{figure}[]
  \centering
   \includegraphics{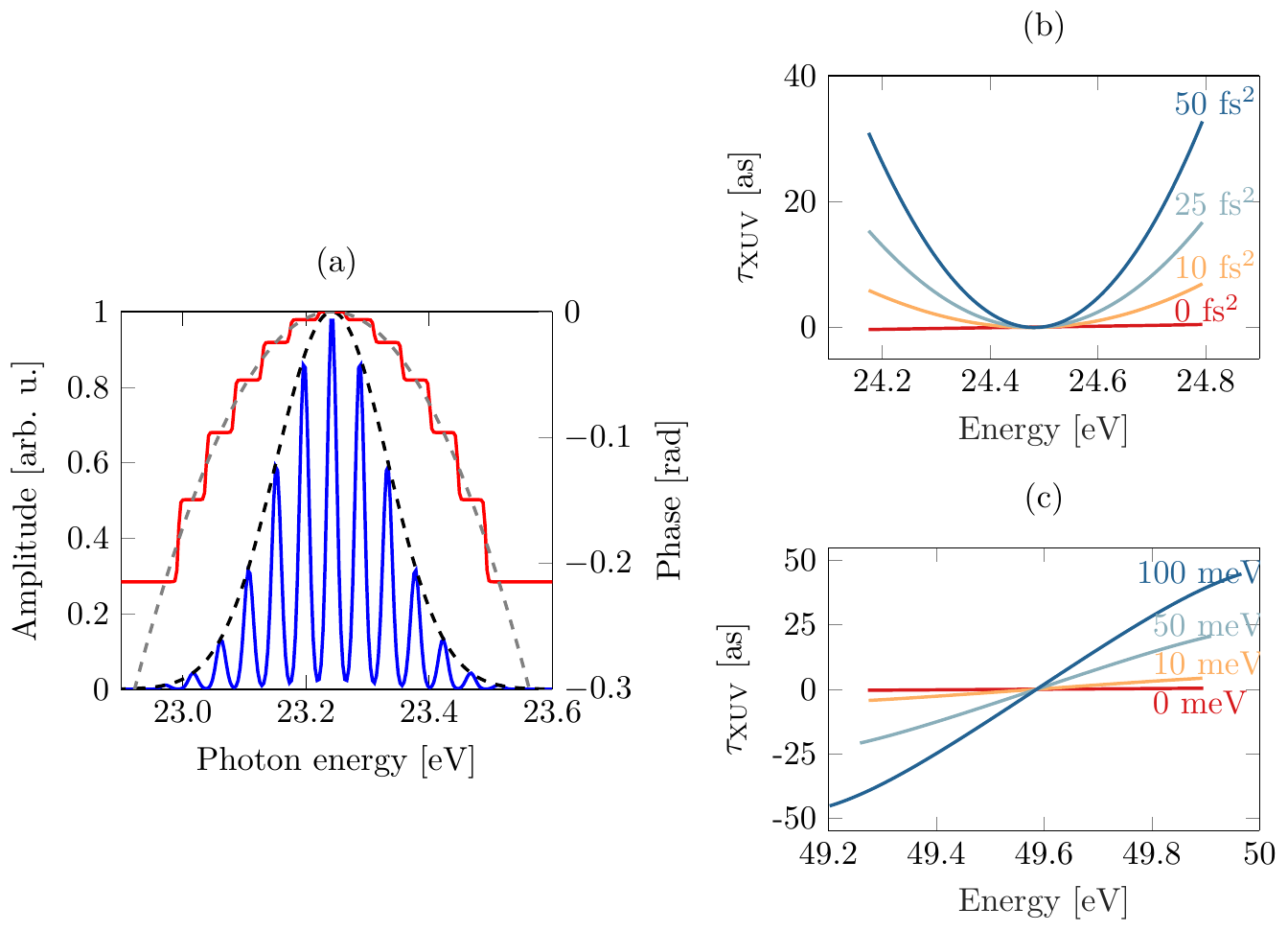}
  \caption{Intra-sideband delay variation. (a) Numerical decomposition of harmonic 15. The dashed black curve is the amplitude of the pulse and the dashed gray curve is the phase. The pulse is decomposed in a 13 different sub-pulses, each with a different CEP. The blue curve shows the resulting amplitude and the red curve the resulting phase. (b) Delay variation in sideband 16 for different values of $\Delta\phi_q^{''}$. There is no blue shift. (c) Delay variation in sideband 32 for different blueshifts. The femtochirp of both harmonics is the same (-50 fs$^2$). Note that in this case, the different phases do not span the same energy as sidebands are broadened by the blueshift. In (b) and (c) the harmonic bandwidths is set to 150 meV and the probe bandwidth to 70 nm}
  \label{Femtochirp_rainbow}
\end{figure}

As it can be seen in Figure \ref{fig:Femtochirp} a, the chirp coefficient has a monotonic decreases as the harmonic order increases.
 This implies that the difference in time between attosecond pulses is not exactly equal to $T/2$.  In a traditional {\it RABBITT} experiment, this effect introduces correction terms to Eq.~(\ref{rabbit2}) and in particular a dependence on the delay due to the difference in chirp-rates  between consecutive harmonics \cite{MauritssonJPB2005}. While this difference in chirp-rates can be considered negligible for long pulses, it becomes important for short pulses.
 To evaluate the error made in the phase retrieval using Eq.~(\ref{rabbit2}) as a function of the fundamental pulse duration, we performed Strong Field Approximation (SFA) simulations  of {\it RABBITT} scans in helium. The results of the simulations are presented in Figure \ref{fig:SFA_chirprate}. Three regions can be identified below 10 fs, between 10-30 fs and above 30 fs.
 At the shortest pulses, large errors are observed due to the almost no periodicity between the minima. Note that an algorithm which also fit the auto-correlation envelope of the sideband signal returned  the same results, ensuring us that the effect at short pulse lengths is not related to the fit model. The error decreases until it reaches a plateau for pulse duration above 30 fs. The height of the plateau of about 5 as reflects the inherent error made by not including the chirp-rate in Eq.~( \ref{rabbit2}).
 The latter is essentially removed when performing two measurements simultaneously. Finally, it is worth mentioning that all correction terms to Eq.~(\ref{rabbit2}) depend on the chirp rate and are therefore sensitive to intensity fluctuations of the fundamental laser. We estimated with SFA simulations that 25\% intensity fluctuations of the fundamental laser can degrade the temporal resolution up to a factor two. The effect of these intensity fluctuations cannot be completely removed by performing two measurements at the same time.
We examine now the influence of the femtosecond chirp on the phase retrieval in the {\it rainbow RABBITT} method. Since the sidebands result from the interference of multiple combinations of IR and XUV frequencies, the spectral characteristics of the light fields becomes important. A rigorous theoretical treatment of {\it rainbow RABBITT} requires a model that goes beyond the approximation presented so far. Here, we adapt the analytic two-photon finite-pulse model developed by Jim\'{e}nez-Gal\'{a}n and co-workers \cite{AlvaroPRA2016} to study the influence of chirped XUV pulses on the retrieved phase from a {\it rainbow RABBITT} measurement.
Since the original finite-pulse model assumes Fourier limited pulses, we numerically decompose the XUV pulse as a sum of 13 sub-pulses each with a different constant phase as shown in Figure \ref{Femtochirp_rainbow}(a), where the dashed curves correspond to the initial amplitude and phase while the solid curves represent the result of the numerical decomposition.
This decomposition allows us to include the phase variation over the harmonic bandwidth, in particular due to the group delay dispersion, which is strongly dependent on the harmonic order. The two-photon transition amplitudes are then computed independently for each sub-pulse and are coherently added. For the calculation, we consider two consecutive harmonics with a bandwidth of 150 meV and a Fourier limited IR probe pulse with a bandwidth of 70 nm. In the absence of resonant states, the measured delay variations can be mainly attributed to $\tau_{\textsc{XUV}}$.

Two different cases are presented based on the results of Fig \ref{fig:Femtochirp} b. The first one, in figure \ref{Femtochirp_rainbow} (b), is for low order harmonics ($q<q_p+10$), where the variation of the GDD as a function of harmonic order, $\Delta\phi^{''}_q=\phi^{''}_{q+1}-\phi^{''}_{q-1}$, can be important. The second one, in Figure \ref{Femtochirp_rainbow} (c), is for high harmonic orders ($q>q_p+10$), where $\Delta \phi^{''}_q$ varies slowly as a function of the harmonic order, and can be neglected. In the first case, the measured delay $\tau_\textsc{XUV}$ can exhibit a quadractic behaviour as a function of electron energy $E$ over the sideband bandwidth due to the difference in GDD between the consecutive harmonics. For instance, when $\Delta \phi^{''}_q=50$ fs$^2$, we can see that the center of the sideband is delayed by more than 30 as compared to the low and high energy parts of the sideband. In such cases, this effect has to be taken into account when measuring the atomic delays $\tau_\mathrm{A}$. However, if both harmonics have close GDD, the spectral phases of the two harmonics compensate each other, giving rise to a constant $\tau_{\textsc{XUV}}$ over the sideband bandwidth, which is a common valid assumption for high harmonic orders.

However, even in the case where both harmonics have the same GDD, the delay may vary over the sideband bandwidth if the contributions from the higher and lower harmonics to the sideband do not perfectly overlap spectrally. Indeed, the IR pulses used to generate harmonics are intense enough so that the front of the pulse can partially ionize the medium through which the rest of the pulse propagates, causing a blue shift of the IR frequency used to generate the harmonics \cite{WahlstromPRA1993}. In most experimental settings, the probe pulse does not propagate through the plasma. The contributions from the lower and higher harmonics will therefore be shifted with respect to each other. This effect has been discussed in \cite{BustoJPB2018} and shown to introduce a linear phase variation which is proportional to the blueshift. In figure \ref{Femtochirp_rainbow}(c), we show the influence of the laser blue shift on the retrieved delay. The GDD of the two neighbouring harmonics is the same (-50 fs$^2$) but the fundamental frequency of the laser pulse used to generate the harmonics is blueshifted while keeping the central frequency of the probe pulse at $\hbar\omega=1.55$  ~eV. A blueshift of the central frequency of the generating IR pulse leads to a non optimal overlap of the contributions of the consecutive harmonics which results in a quasi linear phase variation over the sideband, in qualitative agreement with the predictions of  \cite{BustoJPB2018}. However, the delay extracted from the energy integrated sideband signal will not be affected by this linear phase variation. In general, both GDD variation and blueshift of the fundamental contribute to the total variation of $\tau_\textsc{XUV}$ over the sideband. The magnitude of these effects increases with the duration of both IR and harmonic fields. It is therefore important to monitor the generation conditions to be able to take into account these effects when retrieving the atomic delays.

\section{Optimization of {\it RABBITT} measurements}

Finally, we discuss the influence of fluctuations and statistics on {\it RABBITT} measurements in order to optimize the experimental settings (e.g. step size for sampling equation \ref{rabbit} vs the number of sampled periods).

\subsection{Influence of the temporal jitter}

\begin{figure}[htbp]
   \includegraphics{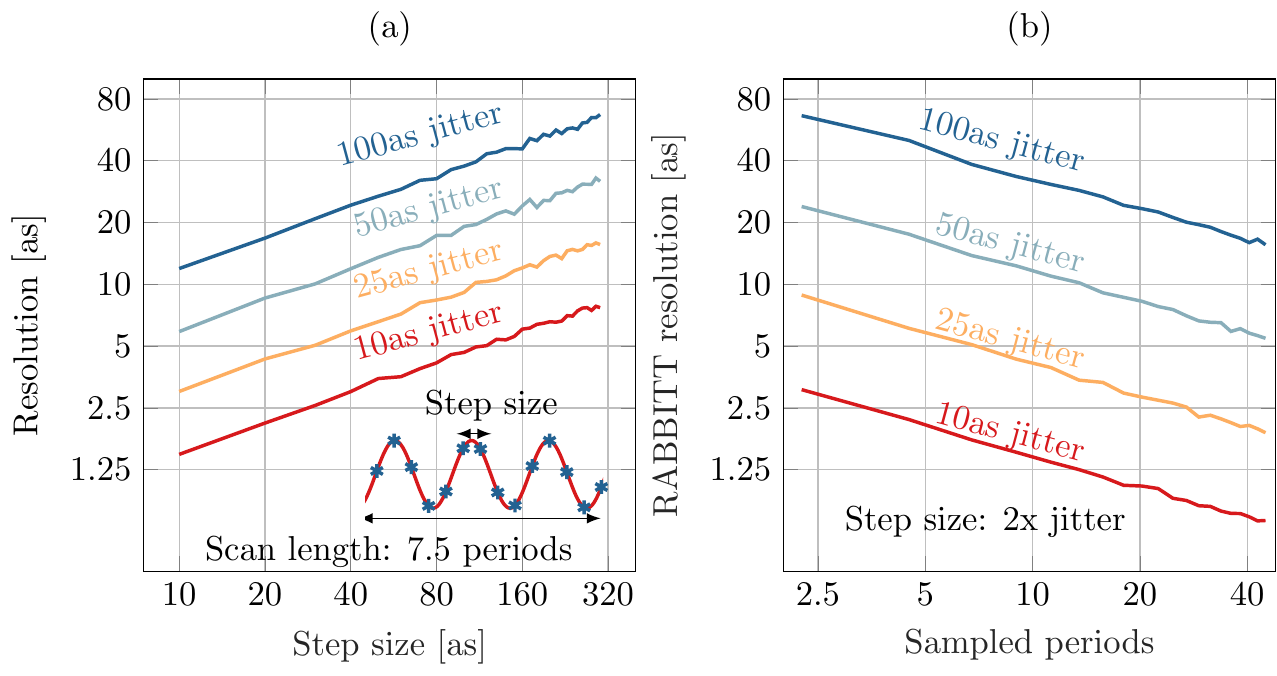}
  \caption{{\it RABBITT} resolution extracted from different sets of 1000 Monte-Carlo simulations. (a) The temporal jitter was set to the four values indicated in the figure and the resolution was calculated while the step size was decreased. (b) The temporal jitter was set to the four values indicated in the figure, and the sampling distance was set to two times the magnitude of the jitter. The resolution was calculated while the number of periods was increased. One period in this case was set to 1.33 fs (i.e. 800 nm wavelength).}
  \label{monte}
\end{figure}

In a {\it RABBITT} measurement, the temporal jitter arises due to a combination of air flow affecting the optical path in each arm and thermal expansion and vibrations of the mirrors in the arms of the interferometer, which cannot be accounted for by the active stabilization loop. Here we evaluate how the temporal jitter affects the accuracy of the retrieved phase, by running a set of Monte-Carlo simulations as described in Section 2d. The parameter space investigated in our simulations consists of the {\it sampling step size}, number of {\it sampled periods} and interferometric {\it stability}, which was set to different values during the simulations. If we neglect any effects related to the femtochirp, we need to sample equation \eqref{rabbit2} with a finite sampling rate and a finite number of periods. An error is immediately introduced due to the finite sampling, we therefore fixed two parameters and analyzed the standard deviation of the error (i.e. the resolution) as a function of the third parameter.

The step size was first varied between 300 as (i.e. just about 4 points per cycle) and 10 as (over 130 points per cycle) for four different jitter amplitudes, ranging from 100 as (the RMS of a good passive interferometer) to 10 as (an extremely good active interferometer). The resulting resolution, as defined in section 2d, is shown in figure \ref{monte} (a). The step size was then fixed and the number of sampled periods was varied between 2 and 40 periods, shown in figure \ref{monte} (b), for the same jitter amplitudes. Both figures \ref{monte} (a-b) exhibit the same trend: if we want to increase by twice the resolution, we need to either decrease the step size by 4 times or sample 4 times as many periods. This is not surprising, since we could have repeated the measurement 4 times instead, decreasing the standard deviation of the mean by a factor $\sqrt{4}$. As mentioned earlier in section 2d, what actually matters is the {\it number} of points with which we sample the sideband, not how we distribute them. Interestingly, even if we oversample, i.e. we sample with a step size smaller than the magnitude of the temporal jitter, we still increase the resolution at the same pace. However, physical limitations in the piezo-controlled delay stage and feedback loop means that we cannot decrease the step size indefinitely. When the physical step size limitation is hit, we can still sample more periods to further increase the resolution. This also has the benefit of increasing the frequency resolution in the Fourier analysis. Since the highest frequency of interest is around $2\omega$, frequencies resolved above it are not usually of interest unless higher multi-photon processes are studied. In general, we found it is optimal in terms of absolute resolution and acquisition time to sample at least 10 periods with a step size of 100 as or less, after which the absolute gain becomes very slow.

\subsection{Influence of statistics}

Here, the influence of the statistical noise and background fluctuations is evaluated. For each sampling point, we acquire a finite number of electrons that make up the sideband signal at each delay step. An error due to statistical noise (fluctuations in detecting electrons) is introduced from a Poisson distribution $P(\sigma^2_s=S(\tau))$.

\begin{figure}[htbp]
   \includegraphics{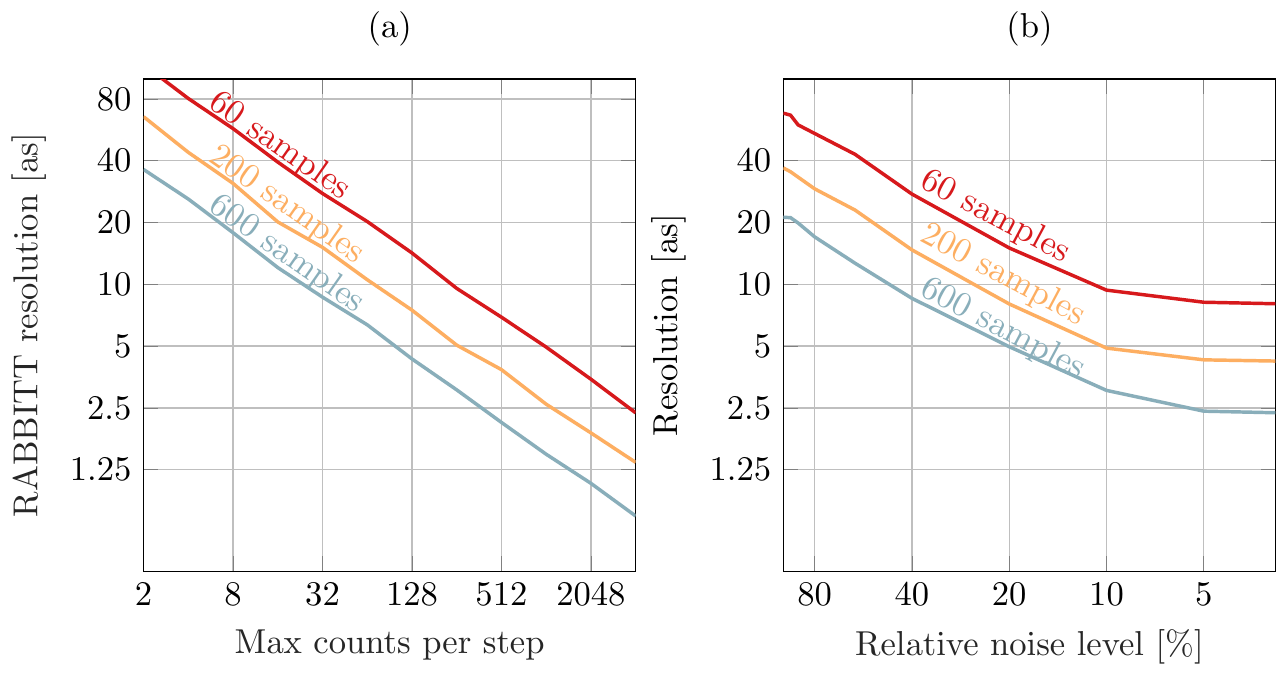}
  \caption{RABITT resolution extracted from a different sets of 1000 Monte-Carlo simulations. (a) The maximum sideband amplitude was varied between 2 and 4086, and the Poisson noise was added to simulate a low the uncertainty of counting electrons in each delay step of the spectrogram. The simulations were repeated for a sample count of 60, 200 and 600 samples with a step size smaller than the Nyquist threshold for $2\omega$. (b) Resolution as a function of relative background noise. The maximum sideband amplitude was set to 500 counts and the noise level was varied between 100\% and 2\% for the same three cases of sample points. }
  \label{monte_2}
\end{figure}

In figure \ref{monte_2} (a), the resolution is studied by varying the maximum sideband amplitude $S(\tau)$ from 2 to 4096 counts (which means that the sideband oscillates between 0 and the maximum amplitude). No background noise is taken into account yet. The trend is the same as in figures \ref{monte} (a-b), which again is expected since we would decrease the standard deviation by a factor of $\frac{1}{\sqrt{N}}$, where N is the number of measurements.

However, once background noise is added the resolution is strongly influenced by the level of noise. In figure \ref{monte_2} (b), the sideband was set to oscillate between 0 and 500 counts, and the background noise was simulated as a white noise, drawn from a uniform distribution between 0\% to 100\% relative noise level. The resolution improves at the same rate as the relative noise level decreases, up to the mark of 10\% relative noise. This is where the resolution reaches the background free limit, see the 500 counts mark in figure \ref{monte_2}(a). These simulations point out that highest resolution can be achieved either by acquiring data with signal-to-noise ratio of $\sim$10 or by filtering the background noise by Fourier analysis.

\subsection{Realistic RABBITT measurement}

\begin{figure}[t]
\centering
   \includegraphics{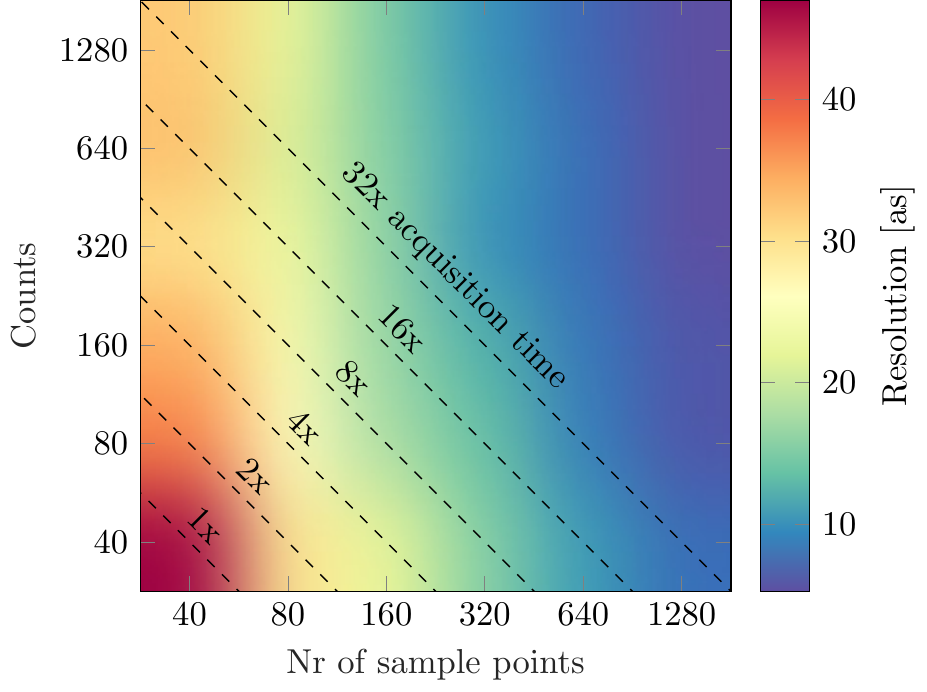}
  \caption{RABBITT resolution as a function of number of sample points and maximum allowed sideband oscillation in measured in counts. Each diagonal line represents a line of constant acquisition time in a real experiment. The simulations were run with a jitter of 50 as and a relative background noise level of 10\%.}
  \label{p_v_c}
\end{figure}

In a real experiment, we have an acquisition time to account for as well. A RABBITT spectrogram consists of many photoelectron spectra taken at different pump-probe delays, meaning the total acquisition time is the acquisition duration of each spectra times the number of delay steps. Hence, we have to weigh the total number of sample points against the time spent at each sample point, which affects the number of acquired counts in a single spectrum. To pinpoint the optimal conditions given a certain RMS of the pump-probe delay jitter, we ran the same Monte Carlo-simulations as above, mapping the two-dimensional parameter space of {\it acquired} counts and {\it number of sample points} for a jitter RMS of 50 as and 10 \% relative background noise. The result is seen in figure \ref{p_v_c}, where the resolution is indicated by the colormap. As the scales are logarithmic, the diagonal lines correspond to a constant acquisition time, i.e. if we spend double the amount of time on acquiring counts on each step, we only sample half the number of oscillation cycles. Each point in the figure then corresponds to a RABBITT measurement. As shown, it is advantageous to spend the duration of the measurement on primarily sampling as many points of the sideband oscillation as possible, and only secondarily to acquire more counts at each sample point. For a certain acquisition time, of which the absolute value naturally depends on the count rate of the experiment, the resolution can vary as much as 3 times depending on if counts or points are prioritized (see the line labelled "32x acquisition time"). As indicated in figure \ref{monte}, a 2x reduction in jitter RMS increases the resolution by the same amount at no acquisition duration cost. The best way to increase resolution, but certainly the most challenging, is to improve the stability of the interferometer.

\section{Conclusions}

In this work, we performed a comprehensive analysis of the {\it RABBITT} technique by investigating the influence of the properties of the light fields and discussing the statistical effects inherent to any measurement. We pointed out two predominant factors often limiting the temporal accuracy to 5-10 as. The femtosecond chirp influences the phase retrieval and should be controlled and accounted for as much as possible. The temporal stability of the controlled delay between the XUV- and IR-fields by the interferometer can be a severe limitation without active stabilization loop.

Monte-Carlo simulations offer a solid tool to devise optimal strategies for data recording depending on the available acquisition time. The simulations show that the total number of sample points of the sideband oscillation determines the accuracy of the phase retrieval, as long as the sampling rate is high enough to resolve the $2\omega$ oscillation. Choosing a step size smaller than the magnitude of the temporal jitter, but higher than the physical limitation of the piezo-actuator, increases the fit accuracy. Fourier analysis can be used to filter data, which can be of great importance if the background noise is above 20\%. Below 20\%, the background noise has a negligible effect. Optimal resolution is found for the highest number of sample points, not the highest number of counts.

To push the temporal resolution below 5 as, further studies and developments should be done. The simple interference equation (Eq.~\ref{rabbit2}) should be extended to include the femtosecond chirp, in order to account for the intrinsic variation in the periodicity. The fundamental laser intensity ought to be monitored in order to filter data. In addition, further studies of the influence of the pump and probe geometries in the interaction volume, including possible spatio-temporal couplings, are required.

\section*{Acknowledgments}
This research was supported by the European Research Council (Advanced grant PALP), the Swedish Research Council and the Knut and Alice Wallenberg Foundation. M.G. acknowledges D. Dowek and the support from the French National Research Agency (ANR) through the “Laboratoire d’Excellence Physics Atoms Light Matter” (ANR10-LABX-0039-PALM).

\bibliographystyle{unsrt}
\bibliography{rsta_bib.bib}

\end{document}